\documentstyle[12pt]{article}
\def \be{\begin{equation}}
\def \ee{\end{equation}}
\def \bea{\begin{eqnarray}}
\def \eea{\end{eqnarray}}

\begin{document}

\begin{titlepage}

\title{\bf The Massive Schwinger Model \\
- a Hamiltonian Lattice Study in a Fast Moving Frame }

\author{
Helmut Kr\"oger and Norbert Scheu \\
\\
\it D\'epartement de Physique,
Universit\'e  Laval, \\
\it Qu\'ebec, Qu\'ebec G1K 7P4,
Canada, \\
\it Email: hkroger@phy.ulaval.ca, nscheu@phy.ulaval.ca
 }

\maketitle

\begin{abstract}
We present a non-perturbative study of the massive Schwinger model.
We use a Hamiltonian approach, based on a momentum lattice corresponding to a
fast moving reference frame, and equal time quantization.
We present numerical results for the mass spectrum of the vector and scalar
particle. We find good agreement with chiral perturbation theory in the strong
coupling regime and also with other non-perturbative studies (Hamer et al., Mo
and Perry) in the non-relativistic regime.
The most important new result is the study of the $\theta$-action, and
computation of vector and scalar masses as a function of the $\theta$-angle.
We find excellent agreement with chiral perturbation theory.
Finally, we give results for the distribution functions.
We compare our results with Bergknoff's variational study from the infinite
momentum frame in the chiral region.
\end{abstract}

\end{titlepage}

\newpage
\section{Introduction}
In order to study $QCD$ non-perturbatively, lattice gauge theory has been most
successful. However, there are some observables, where computational progress
in lattice gauge theory has been slow.
Examples are: Higher excited states of hadrons and mesons, finite density
thermodynamics (quark-gloun plasma), dynamical scattering calculations of cross
sections or phase-shifts and hadron structure functions, in particular in the
region of small $Q^{2}$ and small $x_{B}$ ($10^{-2}$ to $10^{-5}$).
In our opinion a non-perturbative Hamiltonian approach may be a viable
alternative. In deep inelastic lepton-hadron scattering, the success of the
parton model suggests the physical idea to use a fast moving frame also for a
computational study of those processes.
The parton model can be justified using the operator product expansion.
In equal-time quantization, the Breit-frame is the most convenient
choice of frame in order to interpret structure functions.
In Ref.\cite{Kroger97} the authors have proposed such a scheme based on
equal-time quantization, using a lattice Hamiltonian on a momentum lattice
corresponding to a fast moving frame (Breit-frame).
It has been applied to the scalar $\phi^{4}$ theory in $3+1$ dimensions.
Distribution functions and the mass spectrum in the close neighbourhood of the
critical line of the second order phase transition have been computed.

\bigskip

\noindent Here we study $QED_{1+1}$, the so-called massive Schwinger model,
in this framework. This model is physically interesting due to the following
properties:
(i) It is super-renormalizable.
(ii) It has a limit where the model is analytically soluble: the massless
limit~\cite{Schwinger62}.
(iii) It has an (axial) anomaly.
(iv) One can study the topological properties associated with $\theta$-actions
and $\theta$-vacua.
Most of the analytic work on this model has been presented in
Refs.\cite{Lowenstein71}-\cite{Manton85}.
This model is a widely studied benchmark problem for new methods. It has been
studied by a variational method in the infinite-momentum frame
\cite{Bergknoff77},
in discretized light-cone quantization \cite{Eller87}, on the light-cone using
orthogonal functions \cite{Mo93}, on a finite lattice in equal time
quantization
\cite{Crewther80} and by strong coupling series expansion \cite{Hamer97}.

\bigskip

\noindent The purpose of this work is to show:
(a) The use of a fast moving frame, such that $v < c$, in conjunction with {\em
equal-time} quantization works well also in
$QED_{1+1}$. It is not necessary to go to the infinite momentum frame or to
quantize on the light-cone. We obtain quite precise results in the
ultra-relativistic region
$m/g \to 0$.
It has been claimed that low energy states of the massive Schwinger model are
not well aproximated in equal-time field theory. We show here that the problem
is not equal-time quantization, but the use of the rest-frame, which we avoid
to do here.
(b) We consider as most important new results of this work the non-perturbative
results
of the dependence of the vector and scalar mass on the $\theta$-angle.

\section{Method}
Starting from the Lagrangian, we use the axial gauge $A^{3}=0$ to obtain the
Hamiltonian
\be
H = \int_{-L}^{L} dx^{3}
(\bar{\psi} \gamma^{3} i \partial_{3} \psi + \bar{\psi} \psi )
+ \frac{g^{2}}{2} \int_{-L}^{L} dx^{3}
(\psi^{\dagger}\psi) \frac{1}{-\partial^{2}_{3} } (\psi^{\dagger}\psi).
\ee
One introduces a space-time lattice given by spacing $a$ with
$N=\frac{2L}{a}$ lattice nodes. Via discrete Fourier transformation one goes
over to a momentum lattice, with cut-off $\Lambda = { \pi \over a}$ and
resolution
$\Delta p = { \pi \over L}$. Motivated by the parton picture we make the
assumption that left-moving particles are not dynamically important, if
physical particles are considered from a reference frame characterized by a
velocity
$v= { P \over E }$
which is close to the velocity of light. We thus consider a momentum lattice
where $0 \leq p^{0}, p^{3} \leq P$.
In order to minimize the number of virtual particle pairs created
from the vacuum, we choose a small lattice size. The reason for this
is that the number of vacuum pairs is roughly proportional
to a vacuum density times the lattice size.
On the other hand, a fast-moving physical particle is Lorentz contracted;
thus it fits in a small lattice volume (when compared to the rest-frame).

For the purpose of computing the mass spectrum, we need to determine
the vacuum energy. Because the vacuum has the quantum number $\vec P=0$,
the vacuum energy (and only this) is computed in the rest frame.
Because the model is super-renormalizable, one can perform the continuum limit
$a \to 0$. The only renormalization necessary is the subtraction of the vacuum
energy. On a space-time lattice, one has to satisfy a physical condition
(scaling window)  $a < \xi a < L$, where $\xi$ is the correlation length in
dimensionless units, related to the physical mass of the ground state by $M = {
1 \over \xi a}$.
In a strongly relativistic system, $M \ll P$,
thus when $a \to 0$ the scaling window is replaced by
$1/P < 1/M < L$.
For more details compare with Ref.\cite{Kroger97}.

\section{Numerical results}
\subsection{Mass spectrum}
We diagonalize the Hamiltonian in a sector with momentum $\vec{P} = 0$ to
obtain the vacuum energy $E_{vac}$.
Then we diagonalize the Hamiltonian in a sector $\vec{P} \neq 0$ corresponding
to a relativistic velocity. This yields an energy spectrum $E'_{n}$.
The physical energies are obtained from $E_{n} = E'_{n} - E_{vac}$. The mass
spectrum is then given by $M_{n} = \sqrt{ E^{2}_{n} - \vec{P}^{2} }$.
The low lying states of the massive Schwinger model are a vector state and next
a scalar state \cite{Mo93,Hamer97}.
The behaviour in the ultra-relativistic region $ m / g \to 0$ has been computed
in chiral perturbation theory up to second order by Adam \cite{Adam96} and Vary
et al. \cite{Vary96}. The vector particle mass behaves as
\bea
&& \left( \frac{ M_{v} } { M^{(0)} } \right)^{2} = 1 + A \cos(\theta)
{ m \over M^{(0)} } +
[ B - C \cos(2 \theta) ] ({ m \over M^{(0)} })^{2}
+ O( ( { m \over M^{(0)} } )^{3}),
\nonumber \\
&& A = 2 e^{\gamma} \approx 3.5621, ~~~
B \approx 5.4807, ~~~ C \approx 2.0933,
\label{vectormass}
\eea
where $M^{(0)} = g/\sqrt{\pi}$ is the boson mass of the massless Schwinger
model. The mass of the vector boson in the chiral region is shown in Fig.[1].
The vector particle is almost entirely a
fermion-antifermion ($q\bar{q}$) bound state.
In our lattice results presented in Fig.[1] only this $q\bar{q}$ sector has
been taken into account.
We find good agreement with chiral perturbation theory \cite{Adam96,Vary96} and
with finite lattice results by Hamer et al.\cite{Hamer97}.  Fitting our data to
a second order polynomial in $m/g$ yields the coefficients
$A=0.56$, $B=1.74$ and $C=0.22$.
Results for the binding energy $M_{v}-2m$ of the vector boson in the
non-relativistic region are shown in Fig.[2]. We compare our results with
finite lattice results \cite{Hamer97} and with results by Mo and Perry
\cite{Mo93} using light-cone quantization with an expansion in orthogonal
functions.
The corresponding results for the binding energy $M_{s}-2m$ of the scalar boson
are shown in Fig.[3]. Our results correspond to the $q\bar{q}$ sector.
For small masses, i.e. $m/g \approx 1/2$ to $1/8$, the $qq\bar{q}\bar{q}$
sector becomes important.
Thus our results have a larger error in that region.
Results on the scalar mass  in the ultra-relativistic region $m/g <<1 $ taking
into account the $qq\bar{q}\bar{q}$ sector are shown in Figs.[4,5].

\subsection{Dependence on $\theta$-angle}
The massive Schwinger model has $\theta$-vacua and one can study its
$\theta$-action. The $\theta$-physics of the massive Schwinger model on a
circle has been discussed by Manton \cite{Manton85}. The wavefunction is
invariant under local gauge transformations. The $\theta$-angle is a parameter,
which characterizes the behavior of the wavefunction under global gauge
transformations $\Psi[A] \to e^{i n \theta} \Psi[A]$, where $n=0,\pm 1, \pm
2,\cdots$. One can also transform Hilbert space such that this common phase
factor disappears and the wavefunction becomes completely invariant under local
as well as global gauge transformations. This corresponds to introducing a
$\theta$-action $S[A,\theta] = S[A] - \theta q[A]$, where $q[A]$ is the
topological charge, or an equivalent $\theta$-Hamiltonian.
The mass spectrum of low-lying states as a function of the $\theta$-angle is
shown in Figs.[4,5]. Fig.[4] shows results (raw data), where the usual vacuum
energy $E_{vac}(\theta=0)$ has been subtracted in all $\theta$-sectors,
which is unphysical, e.g., for $\theta=\pi$. It shows, however, the behavior of
the vector mass for $\theta=\pi$ when increasing the lattice from $N=6$ to
$24$.
One observes a cusp in the curves, becoming more pronounced when $N$ increases.
This behavior is due to level crossings with higher lying states (not shown in
the Fig.). Fig.[5] shows the same results when the vacuum energy
$E_{vac}(\theta)$ has been subtracted in the corresponding $\theta$-sector.
One observes agreement with first order chiral perturbation theory in the
ultra-relativistic regime ($m/g < 0.04$ for $N=6$ and $m/g < 0.01$ for $N=24$).
The slope of those curves is related to the expectation value of the chiral
condensate. The Feynman-Hellmann theorem relates the fermion condensate
$< \bar{\psi} \psi >$ to the derivative of the vacuum energy
with respect to the fermion mass.
Adam \cite{Adam95} has computed the condensate in lowest order of chiral
perturbation theory,
\bea
&& \frac{< \bar{\psi} \psi > }{ M^{(0)} } = a \cos(\theta)
+ [ b - c \cos(2 \theta) ] { m \over M^{(0)} } + O( { m \over M^{(0)} } )^{2}).
\nonumber \\
&& a = { e^{\gamma} \over 2 \pi } \approx 0.2835, ~~~ b\approx 0.7825,
{}~~~c\approx 0.7163.
\label{condensate}
\eea
Eqs.(\ref{vectormass},\ref{condensate}) imply
\be
< \bar{\psi} \psi > = \left. { M^{(0)} \over 2 \pi }
{ \partial \over \partial m } M_{v} \right|_{m=0},
\ee
which in analogy to the Feynman-Hellmann theorem relates the condensate to the
vector mass. When we extract the slope $\partial M_{v} / \partial m$ from
Fig.[5], we obtain $< \bar{\psi} \psi > /g = 0.16 \cos(\theta)$, compared to
the exact result of the massless Schwinger model given by
\be
\frac{ < \bar{\psi} \psi > } { g } = \frac{ e^{\gamma} }{2 \pi \sqrt{\pi} }
\cos(\theta) \approx 0.1599 \cos(\theta).
\ee
The $\theta$-dependence and global gauge
transformations have been treated in standard lattice gauge theory based
on compact link variables $U_{\mu}(x)$ by Hamer et al. \cite{Hamer82}.
Our results are in agreement with theirs.

\subsection{Distribution functions}
In the Hamiltonian approach it is easy to compute the wave function of a
low-lying state.
{}From the wave function one can obtain information on its parton structure,
i.e., the number of partons and their momentum distribution.
The distribution function of the vector boson is given by
\be
f(x_{B}) =
 <\Psi_{v}(P) \mid
   {1\over 2}
   \left[ b^{\dagger}_{k} b_{k}
        + d^{\dagger}_{k} d_{k}
  \right]\mid \Psi_{v}(P) >,
\ee
where $x_{B} = k/P$ is the fraction of momentum of the vector boson carried by
the parton, i.e., fermion. In Fig.[6] we display our lattice results for $m/g =
0$ to $0.28$. In the massless limit
the distribution function has the shape of a box. Our results are compared with
variational calculations using the infinite-momentum frame by Bergknoff
\cite{Bergknoff77} and also with those by Mo and Perry \cite{Mo93} using the
light-cone. The most sensitive region is the ultra-relativistic region. We find
 agreement in shape with Mo and Perry's results and
very good agreement with Bergknoff's results as shown in Fig.[7].
The agreement with Mo and Perry's results becomes even better than
the agreement with Bergknoff if their
distribution function is
normalized to one.
In Fig.[8] we also display the convergence behavior of the vector distribution
function for $m/g=32$, i.e., in the non-relativistic region. It can be seen
that a reasonably large lattice ($N \approx 200$) is needed to resolve this
function (which has a sharp peak at $x_{B}=1/2$
when $m/g \to \infty$).

\section{Summary}
In this work we have applied a Hamiltonian lattice approach in a fast moving
reference frame to study the massive Schwinger model.
We find that the method works well for the computation of the low-lying mass
spectrum, also in the presence of the $\theta$-action, and distribution
functions. Here we have investigated only $\theta=0,\pi$. It is interesting to
further study the dependence on other $\theta$-angles and higher-lying states.
More $\theta$-angles can be studied, e.g. by adding one negative momentum state
to the basis. Those investigations are feasible, but would require a higher
computational effort.

\bigskip

\noindent {\bf Acknowledgement} \\
This work has been supported by NSERC Canada. N.S. acknowledges support by an
AUFE fellowship from DAAD Germany.

\newpage
\begin{flushleft}
{\bf Figure Caption}
\end{flushleft}
\begin{description}

\item[{Fig.1}]
The dimensionless mass of the vector boson versus $m/g$ in the chiral region.
Comparison with Hamer et al. \cite{Hamer97} and chiral perturbation theory
\cite{Adam96,Vary96}.

\item[{Fig.2}]
The dimensionless binding energy of the vector boson versus $\log_{2}(m/g)$ in
the non-relativistic region. Comparison with Hamer et al. \cite{Hamer97} and
Mo and Perry \cite{Mo93}.

\item[{Fig.3}]
Same as Fig.[2], but for scalar boson.

\item[{Fig.4}]
Dependence of the dimensionless mass of vector and scalar particle
on the $\theta$-angle before proper subtraction of $\theta$-vacuum energy (raw
data).
Comparison with chiral perturbation theory \cite{Adam96,Vary96}.

\item[{Fig.5}]
Dependence of the dimensionless mass of vector and scalar particle
on the $\theta$-angle after subtraction of $\theta$-vacuum energy.
Comparison with chiral perturbation theory \cite{Adam96,Vary96}.

\item[{Fig.6}]
Distribution function of the vector boson for different $m/g$ in the chiral
region.

\item[{Fig.7}]
Distribution function of the vector boson for $m/g=0.1/\sqrt{\pi}$.
Comparison with results by Mo and Perry \cite{Mo93} and Bergknoff
\cite{Bergknoff77}.

\item[{Fig.8}]
Distribution function of the vector boson for $m/g=32$ in the non-relativistic
region.

\end{description}

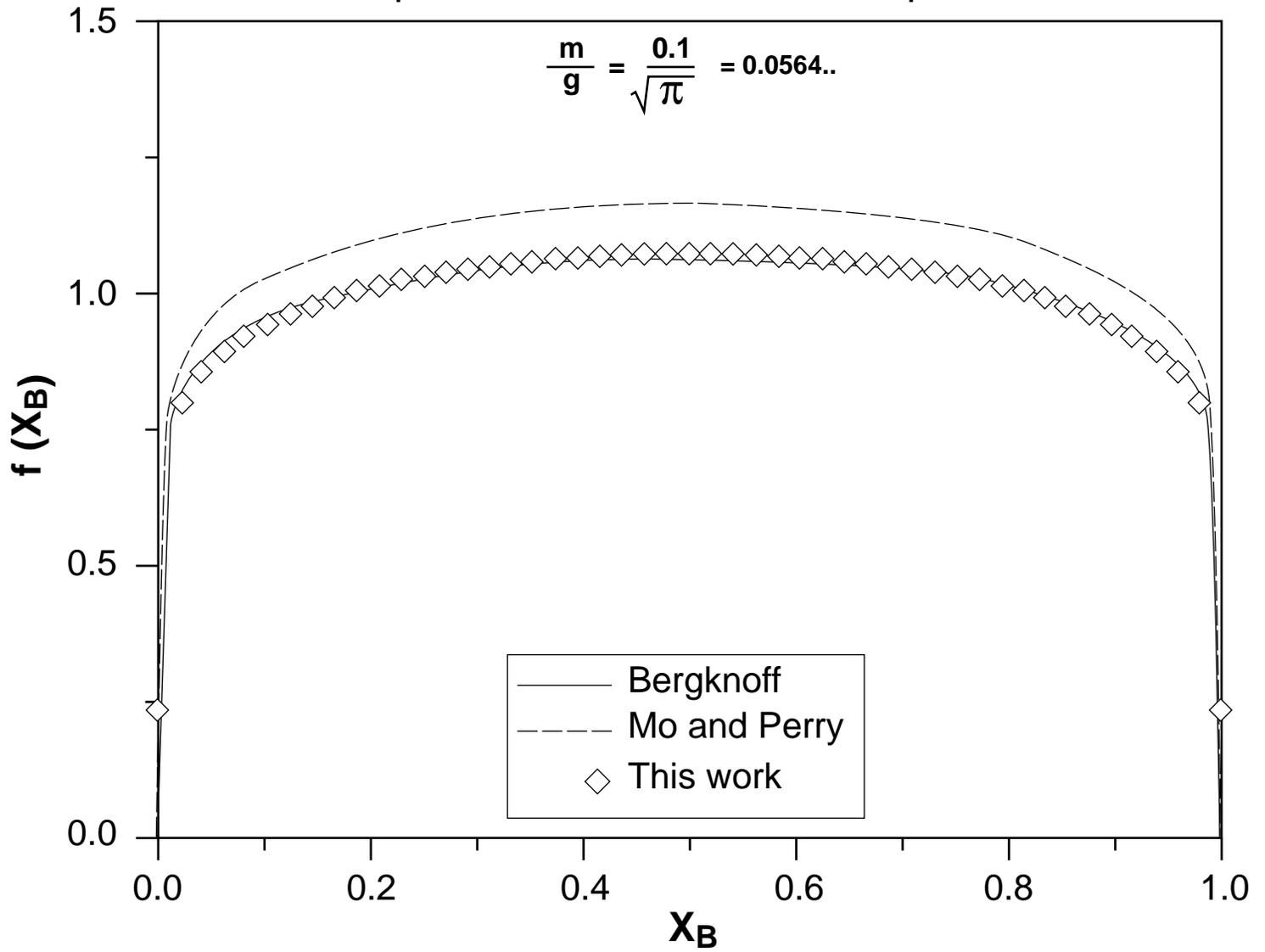
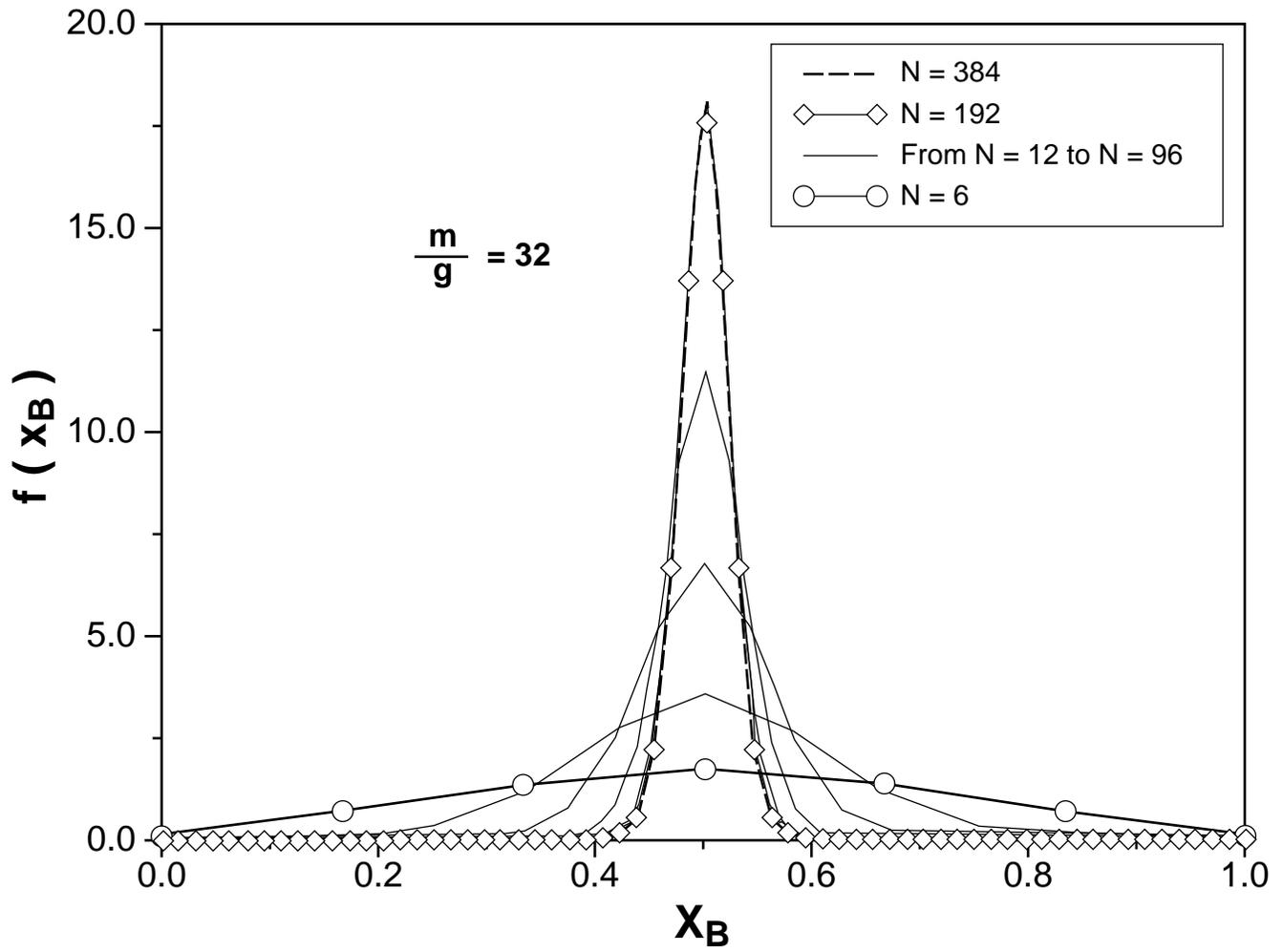
\end{document}